\documentclass[letterpaper,8pt]{extarticle}  
\usepackage{import}
\usepackage{local}
\usepackage[left=.65in,top=.65in,bottom=.65in,right=.65in]{geometry}
\usepackage{lipsum} 

\title{Mathematical models of the \textit{Arabidopsis} circadian oscillator}

\author{%
\textbf{Lucas Henao \orcidlink{0009-0001-8083-0586} ,\textcolor{Accent}{\textsuperscript{1,2}} %
Sa\'ul Ares \orcidlink{0000-0001-6214-4083},\textcolor{Accent}{\textsuperscript{1,3,*}} %
Pablo Catal\'an \orcidlink{0000-0001-6214-4083}\textcolor{Accent}{\textsuperscript{1,2,*}} }\\
\begin{small}\textcolor{Accent}{\textsuperscript{1}} Grupo Interdisciplinar de Sistemas Complejos (GISC), Madrid, Spain \\ 
\textcolor{Accent}{\textsuperscript{2}} Department of Mathematics, Universidad Carlos III de Madrid, 28911 Legan\'es, Spain \\
\textcolor{Accent}{\textsuperscript{3}} Centro Nacional de Biotecnologia (CNB), CSIC, 28049 Madrid, Spain \\
\textcolor{Accent}{\textsuperscript{*}}Correspondence: \textcolor{Accent}{saul.ares@csic.es (SA), pcatalan@math.uc3m.es (PC)} \\ \end{small}
}

\date{}
\begin{document}
\maketitle
\thispagestyle{empty}

\section{Abstract}

\begin{doublespacing}

\noindent
\textbf{\textcolor{Accent}{We review the construction and evolution of mathematical models of the \textit{Arabidopsis} circadian clock, structuring the discussion into two distinct historical phases of modeling strategies: extension and reduction. The extension phase explores the bottom-up assembly of regulatory networks introducing as many components and interactions as possible in order to capture the oscillatory nature of the clock. The reduction phase deals with functional decomposition, distilling complex models to their essential dynamical repertoire. Current challenges in this field, including the integration of spatial considerations and environmental influences like light and temperature, are also discussed. The review emphasizes the ongoing need for models that balance molecular detail with practical simplicity.}}

\section{Introduction}
In biology, explanations are given by mechanisms: \textit{how} and \textit{why} are usually treated as equivalent questions. Mechanisms can be represented through models describing their essential parts and interactions. Moreover, models illustrate how these parts and interactions fit within a system to produce an observed behavior. Consequently, modeling requires decomposition of the system into parts, and identifying those performing the operations giving rise to the phenomenon of interest \cite{bechtelDynamicMechanisticExplanation2010}. This decomposition may proceed in several ways, which will in turn modify our understanding of the system's function \cite{kauffmanArticulationPartsExplanation1970}. 

In systems biology, these working units are commonly network motifs \cite{alonIntroductionSystemsBiology2020} or regulatory modules \cite{jaegerDynamicalModulesMetabolism2021} embedded in gene regulatory networks. The main method to probe into mechanisms in a system thus decomposed is through genetic perturbations \cite{jaegerDynamicalModulesMetabolism2021}. Therefore, modeling proceeds iteratively through a constant dialogue between theory and experiment \cite{rodriguezmaroto2024mathematical}. This iterative cycle establishes a positive feedback loop by which, as more data becomes available, model construction accelerates. The constructed models, in turn, guide further experimental work. Nevertheless, indefinite continuation of this cycle is neither possible nor desirable for practical and computational reasons. As models become increasingly complex and detailed, they also become less tractable, that is, harder to develop further, analyze and manipulate. Arbitrary detail may be obtained by steadily decomposing the system into ever-growing lists of parts, but dynamic mechanistic explanation requires an account of the working parts. 

We focus here on the development of mathematical models of the \textit{Arabidopsis} circadian clock, dividing them in two phases that we have named extension and reduction phases. The extension phase consists on the bottom-up construction of regulatory networks to accurately reproduce the oscillatory behavior of the molecular components of the clock. It is thus concerned with structural decomposition of the system. The main model-lineage is represented by models of ordinary differential equations (ODEs) coming from Andrew Millar's laboratory at the University of Edinburgh. Several of the models in the expansion phase have already been reviewed in \cite{bujdosoMathematicalModelingOscillating2013}. The reduction phase is instead centered on functional decomposition and (to our knowledge) has not been reviewed so far. In the reduction phase, the constructed models are pruned such that the essential regulatory modules are identified and the key dynamical features of the model are studied.

In what follows, we review the principal models of the plant circadian clock, classifying them under either the expansion or reduction phase. Finally, we cover recent advances in the field and discuss three of the main challenges ahead for clock models, namely the inclusion of spatial structure in the model, or considering the roles of either light or temperature on the properties of the clock.





\section{Chronological development of \textit{Arabidopsis} circadian clock models: Expansion phase}
\begin{figure}[!t]
    \centering
    \includegraphics[width=0.8\linewidth]{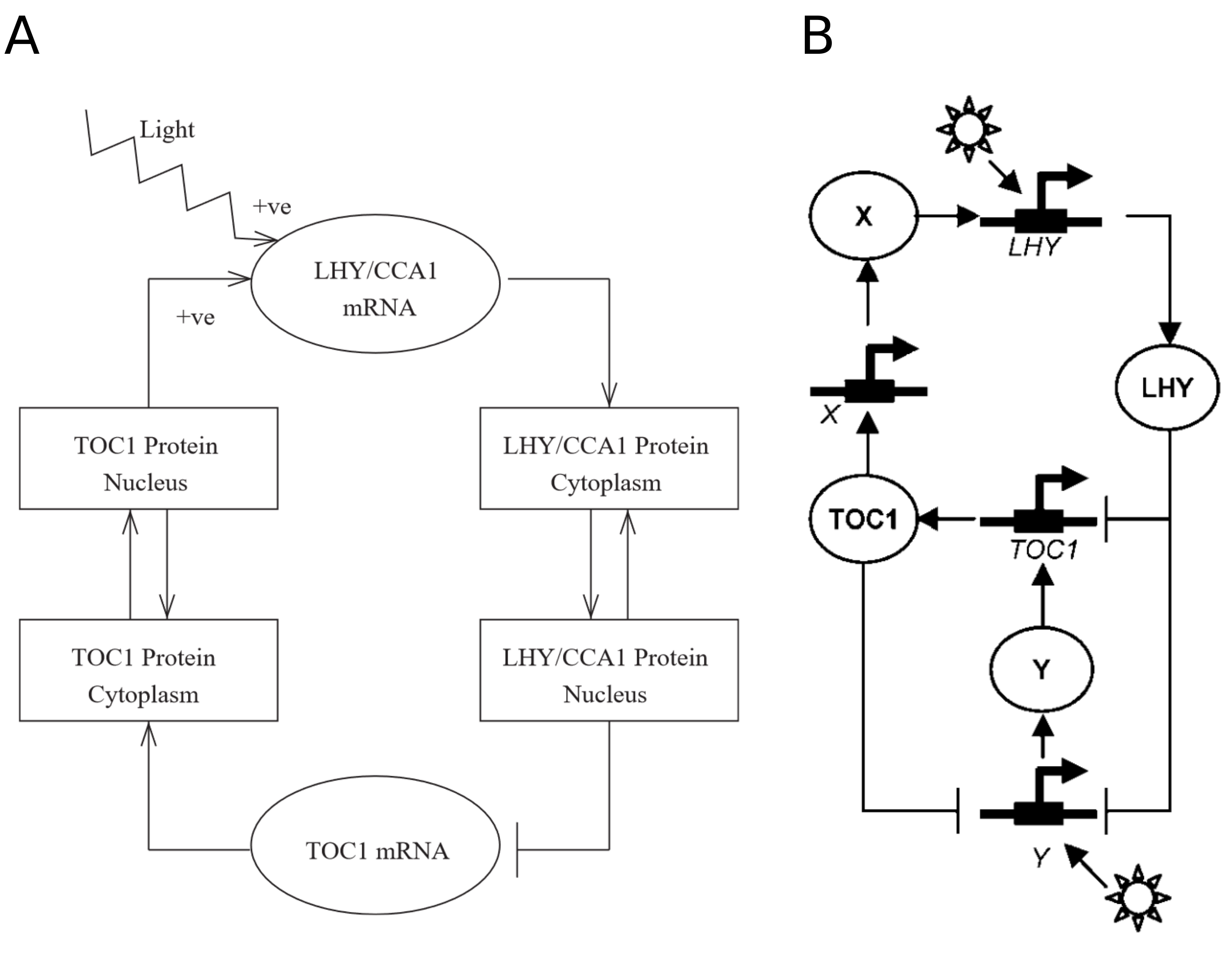}
    \caption{Structure of the circadian clock in L2005a and L2005b. (A) The circadian clock in L2005a consists of a single feedback loop, where LHY/CCA1 inhibit the expression of TOC1, which activates them. (B) Two new unknown components, X and Y, are included in L2005b to account for experimental data. Reproduced with permission from \cite{lockeModellingGeneticNetworks2005} (A), and with modifications from \cite{lockeExtensionGeneticNetwork2005} (B), Creative Commons license.}
    \label{L2005}
\end{figure}

At the roots of \textit{Arabidopsis thaliana} circadian oscillator modeling is Locke's 2005 model, henceforth named L2005a. ({The convention so far with \textit{Arabidopsis} clock models has been to name them using the name of the first author and either the year of submission or publication. In what follows, we use the names of the models given by later works citing them.}) L2005a  \cite{lockeModellingGeneticNetworks2005}, comprised of a single negative feedback loop, the simplest network motif with oscillatory behavior \cite{alonIntroductionSystemsBiology2020}. It is a transcription-translation feedback loop wherein the transcription factors \textit{CIRCADIAN CLOCK ASSOCIATED 1} (CCA1) and \textit{LATE ELONGATED HYPOCOTYL} (LHY), modeled as a single variable because of their partially redundant roles, repress the transcription of \textit{TIMING OF CAB EXPRESSION 1} (TOC1), which positively regulates the former (Fig.~\ref{L2005}A). Entrainment is modeled through light activation on LHY/CCA1 transcription. This makes the network unable to respond to day length, as LHY activation happens only at dawn, rendering the model insensitive to light at the end of the photoperiod.

This system makes for an effective oscillator, but it falls short in predicting most experimental observations. At least two new components are needed: (i) ``X'', to explain the delay between TOC1 transcription in the evening and LHY/CCA1 activation in the following morning; and (ii) ``Y'', an additional light-responsive component to account for LHY/CCA1-independent entrainment. ``X'' mediates LHY activation by TOC1 and ``Y'' forms a feedback loop with TOC1, which explains the oscillatory behavior found in \textit{lhy;cca1} double mutants. These two components are added in a new version of the model, which we call L2005b \cite{lockeExtensionGeneticNetwork2005}. This interlocked feedback loop (Fig.~\ref{L2005}B) is able to account for the experimental data available at the time and shows a robust behavior even upon parameter changes, thus overcoming the weaknesses of the preceding single loop model.

\begin{figure}[!t]
    \centering
    \includegraphics[width=0.6\linewidth]{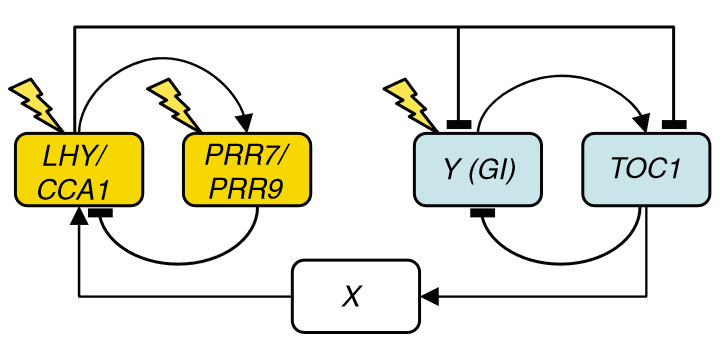}
    \caption{Structure of the circadian clock in L2006. The clock now consists of three loops: a morning loop (yellow), an evening loop (blue) and an overarching loop that connects both, including LHY/CCA1, TOC1 and the unknown component X. Only genes are shown, and arrows represent regulatory interactions. Reproduced from \cite{lockeExperimentalValidationPredicted2006}, Creative Commons license.}
    \label{L2006}
\end{figure}

Two models published simultaneously by Zeilinger \textit{et al.} \cite{zeilingerNovelComputationalModel2006} and Locke \textit{et al.} \cite{lockeExperimentalValidationPredicted2006}, henceforth L2006, independently extended L2005b by adding \textit{PSEUDO-RESPONSE REGULATOR 7} and \textit{9} (PRR7 and PRR9). In L2006, PRR7 and PRR9 are modeled as a single component that represses LHY/CCA1 and is in turn activated by LHY/CCA1. The resulting structure is a three-loop system consisting of two short period oscillators, the morning (PRR7/9-LHY/CCA1) and evening (TOC1-Y) loops, coupled by the LHY/CCA1-TOC1-X loop (Figure \ref{L2006}). Y is experimentally confirmed to be a component of \textit{GIGANTEA} (GI), and it is thought to be the light responsive element of the evening loop, as the \textit{lhy/cca1} mutant still maintains entrainable oscillations and TOC1 is not observed to be light-activated.

\begin{figure}[!b]
    \centering
    \includegraphics[width=0.7\linewidth]{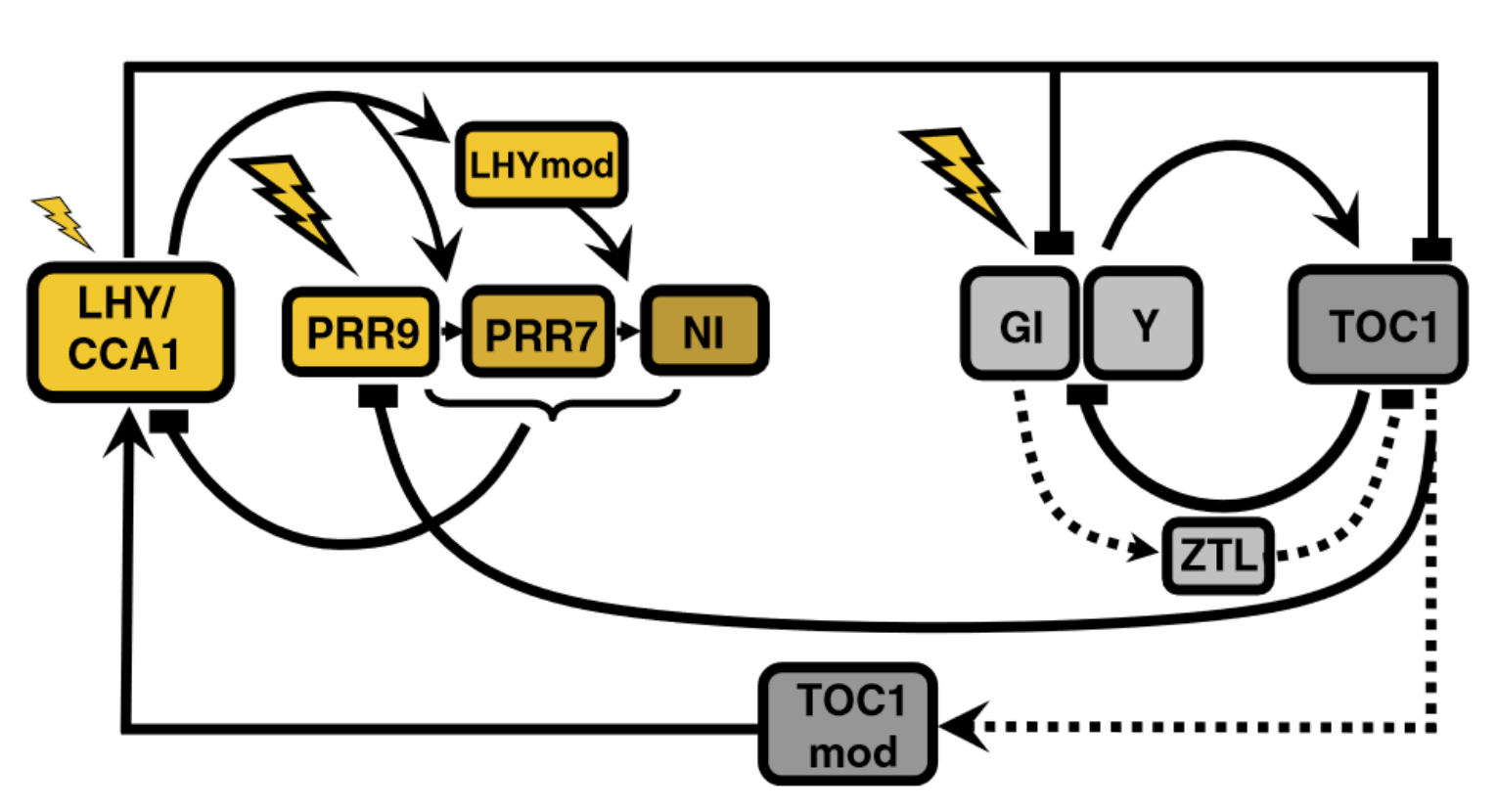}
    \caption{Structure of the circadian clock in P2010. New components enter the model, both in the morning (NI) and evening (ZTL) loops. Post-transcriptional modification of proteins is included for the first time. Unknown component X is substituted by a TOC1-dependent complex. Modified from \cite{pokhilkoDataAssimilationConstrains2010}, Creative Commons license.}
    \label{P2010}
\end{figure}

The models so far described are based on transcriptional control and do not include post-transcriptional regulation. Pokhilko's \textit{et al.} 2010 model \cite{pokhilkoDataAssimilationConstrains2010}, henceforth P2010, is the first model to do so by including \textit{ZEITLUPE} (ZTL)-induced acceleration of TOC1 protein degradation (Fig.~\ref{P2010}). ZTL function had been acknowledged in previous models, but not explicitly modeled.

High TOC1 mRNA levels in \textit{gi} mutants bring into question the assumption in L2006 that GI activates TOC1, prompting the separation of GI and Y functions: ZTL stabilization is done by GI (observed experimentally) and promotion of TOC1 transcription is due to Y. This regulatory scheme makes up the evening loop.

The morning-evening bridge suffers two modifications: the replacement of X by a TOC1-dependent complex and the addition of TOC1 inhibition of PRR9 expression, which explains the experimentally observed reduction of PRR9 mRNA levels when TOC1 is overexpressed. These interactions describe a coherent type 1 feedforward loop (C1-FFL): TOC1 positively regulates LHY both through TOC\textsubscript{mod} (see Fig.~\ref{P2010}), and indirectly via repression of PRR9; LHY then feeds back into TOC1. This network motif buffers noise in input stimuli, acting as a persistence detector \cite{manganCoherentFeedforwardLoop2003}. 

The morning loop is extended by adding a ``night inhibitor'', NI, as part of the inhibitor wave: PRR9, PRR7 and NI act sequentially, ensuring that LHY/CCA1  mRNA is kept at very low levels from dawn until the middle of the night. Then, degradation of the components of this inhibitor wave and activation by TOC\textsubscript{mod} quickly elevates mRNA levels again through the aforementioned C1-FFL. However, the mechanism governing this switch from inhibition to activation is not explained.

In the morning loop, PRR9 is the light-sensitive component, so light initiates the inhibitor wave. In the evening loop, Y becomes the light-sensitive component, positively regulating TOC1. The model also includes stabilization by light of these components, making this wave dawn-responsive too, and thus improving the response to changing photoperiods.

The models discussed until now do not allow arrhythmic clocks when performing a loss-of-function mutation in one gene: they only support speeding-up or slowing-down of the clock. This theoretical premise is challenged by the observation that knocking out the gene \textit{EARLY FLOWERING 4} (ELF4) brings the oscillator to a halt \cite{mcwattersELF4RequiredOscillatory2007}. Following this observation, Kolmos \textit{et al.} \cite{kolmosIntegratingELF4Circadian2009} placed ELF4 in the clock model for the first time in 2009, although not mathematically, and hypothesized that it functions as an integrator of the morning and evening loops ---the missing piece in P2010. 

Arrhythmic mutants exist at another two loci, \textit{EARLY FLOWERING 3} (ELF3) and \textit{LUX ARRHYTHMO} (LUX) \cite{nusinowELF4ELF3LUX2011}. ELF3 and LUX form, together with ELF4, the Evening Complex (EC), which explains the fact that all three mutants have a similar phenotype. Herrero \textit{et al.} \cite{herreroEARLYFLOWERING4Recruitment2012} used Linear Time Invariant (LTI) modeling to integrate ELF4 and ELF3 in the L2006 model. LTI modeling not only confirmed already existing connections, but also predicted ELF3/ELF4 inhibition by LHY and PRR9/PRR7 inhibition by ELF3/ELF4. An important observation is that ELF3 and LUX overexpression restores rhythm in \textit{elf4} mutants, so rhythmic expression of PRR9 requires both CCA1/LHY-mediated transcriptional activation and cooperative transcriptional repression by the components of the EC.


\begin{figure}[!b]
    \centering
    \includegraphics[width=0.9\linewidth]{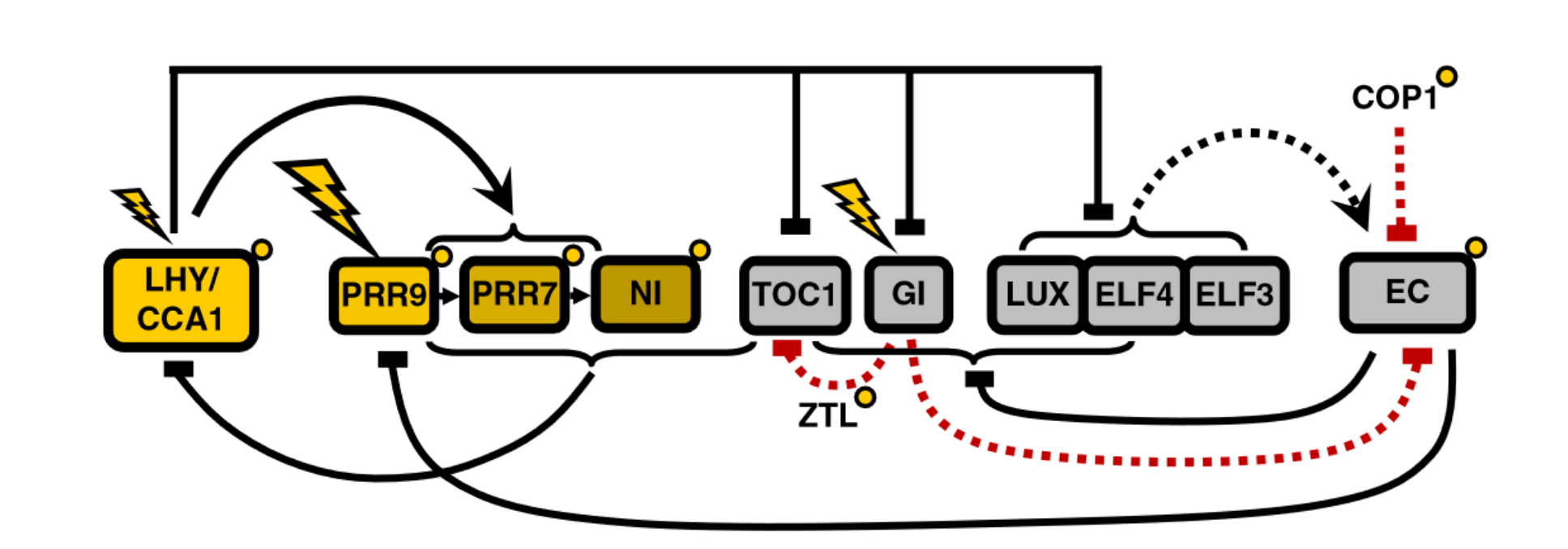}
    \caption{Structure of the circadian clock in P2011. The main novelty is the introduction of the Evening Complex (EC) into the model, substituting the unknown component Y and changing the relationship between the evening and morning loops. Modified from \cite{pokhilkoClockGeneCircuit2012}, Creative Commons license.}
    \label{P2011}
\end{figure}

The clock model undergoes significant modifications in the next iteration of Pokhilko's model, P2011 \cite{pokhilkoClockGeneCircuit2012}, motivated mainly by the inclusion of the EC (Fig.~\ref{P2011}).
This restructuring of the model starts from the evening loop, now composed of the EC and its separate components ---which replace Y. The EC inhibits the expression of both ELF4 and LUX, creating a negative feedback loop, and also represses TOC1. Additionally, the EC is regulated by COP1 \cite{nietoCOP1DynamicsIntegrate2022} and GI in a light-dependent manner: entrainment of the evening loop happens through COP1-mediated degradation of ELF3, and GI acts as a modulator in the evening loop by negatively regulating the EC and thus indirectly activating TOC1 via double inhibition.

\begin{figure}[!b]
    \centering
    \includegraphics[width=0.9\linewidth]{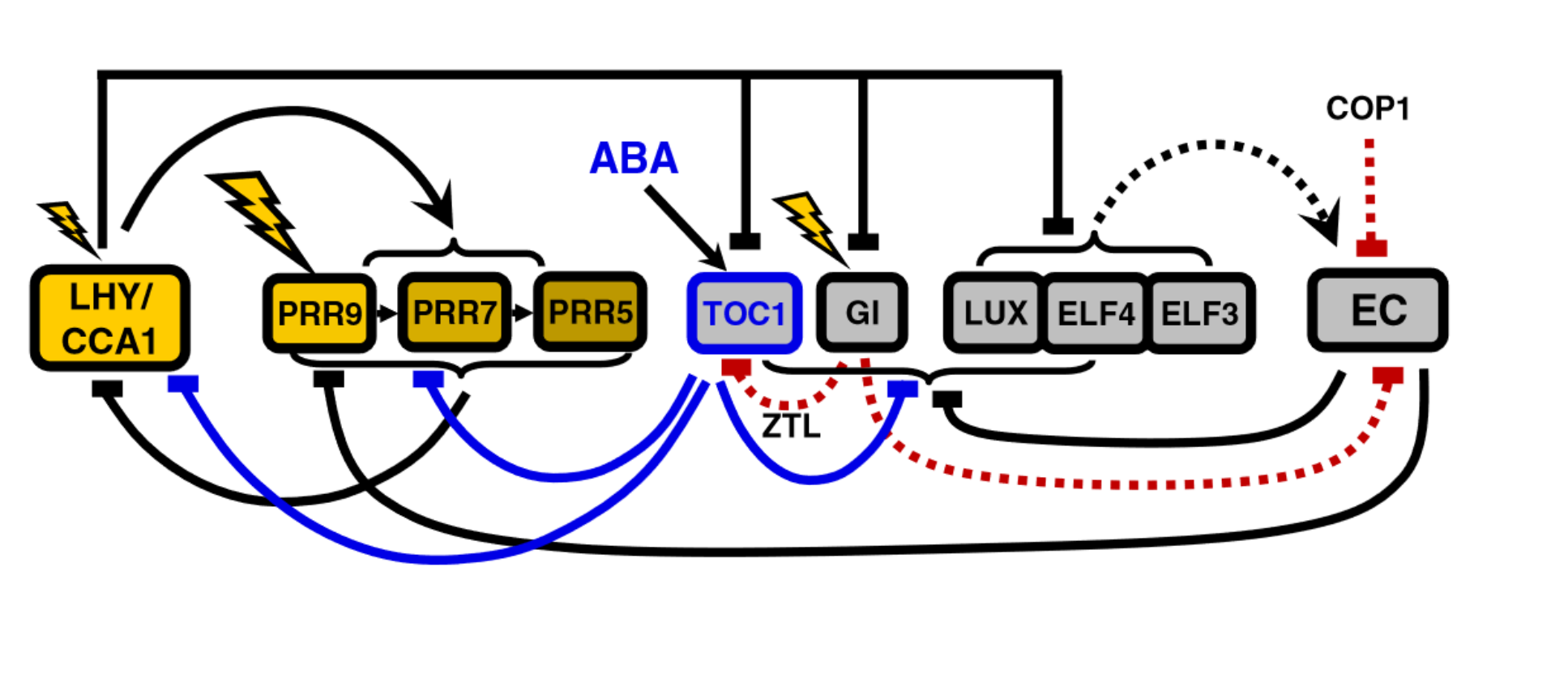}
    \caption{Structure of the circadian clock in P2012. Note the widespread repressive activity of TOC1, which reflects recent experimental data cited in \cite{pokhilkoModellingWidespreadEffects2013}. The model also includes induction of TOC1 transcription by Abscisic Acid (ABA). Modified from \cite{pokhilkoModellingWidespreadEffects2013}, Creative Commons license.}
    \label{P2012}
\end{figure}

The morning loop retains its main features ---the feedback loop between LHY/CCA1 and the inhibitor wave. However, the role of TOC1 as a regulator of the morning loop is revised: it is no longer an activator, but rather a repressor of LHY/CCA1, making it another component of the inhibitor wave. This is consistent with the fact that it also belongs to the PRR gene family. This new role fits better the data from various mutants and eliminates the need for the hypothetical X or TOC\textsubscript{mod}. 

In agreement with experimental findings, modeling of EC mutants shows reduction in the amplitude  of LHY/CCA1 oscillations. The structure of the model suggests that this is due to the higher level of its EC-regulated inhibitors, TOC1 and PRR9. Repression of both of these LHY/CCA1 inhibitors is required for robust LHY/CCA1 oscillations and thus anticipation of dawn by the clock; the morning loop is unable to support self-sustaining oscillations. This underscores the central role of the EC.

Regulation of the evening loop by the morning loop happens through repression of TOC1, LUX, ELF4, ELF3 and GI expression by LHY/CCA1.

As a result of these modifications, the circadian clock now contains a repressilator \cite{elowitz2000synthetic}, consisting of a triad of sequential negative regulation whereby earlier-expressed clock components are repressed by the later ones: LHY/CCA1 represses the EC, which represses the PRRs, which in turn repress LHY/CCA1. The only positive interaction remaining in the model is LHY/CCA1 promotion of the inhibitor wave.

The main theme of the remaining models of the expansion phase is the reinterpretation of direct positive regulation as double inhibition. In this regard, the inhibitory role of TOC1 as a repressor first introduced in P2011 is developed further in the last iteration of Pokhilko's model, P2012 \cite{pokhilkoModellingWidespreadEffects2013}: TOC1 directly inhibits PRR9, PRR7, PRR5, LUX, ELF4 and GI in addition to LHY/CCA1 (Fig.~\ref{P2012}).

Fogelmark and Troein's model, henceforth F2014 \cite{fogelmarkRethinkingTranscriptionalActivation2014}, represents the final model of the expansion phase and is consequently the most intricate of all. The remaining transcriptional activation in previous models is removed, in view of the lack of direct experimental data supporting this role in any component of the clock.
Instead, a novel component is included, the transcription factor \textit{REVEILLE 8} (RVE8), activating the expression of a large portion of clock genes. Another addition is that of the night-expressed transcription factor \textit{NOX/BROTHER OF LUX ARRHYTHMO} (NOX/BOA).

\begin{figure}[!t]
    \centering
    \includegraphics[width=0.9\linewidth]{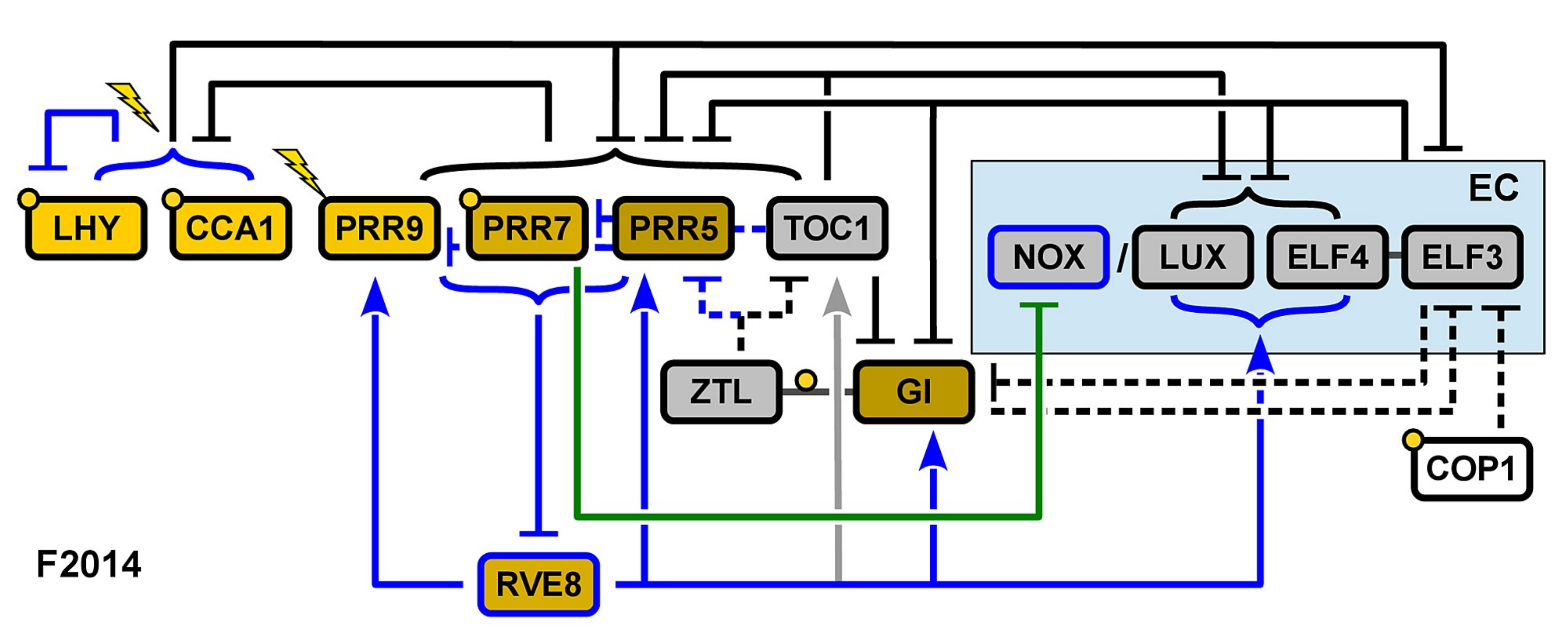}
    \caption{Structure of the circadian clock in F2014. Two new components are included in the model, RVE8 and NOX. Blue arrows indicate new interactions included in the model. Reproduced from \cite{fogelmarkRethinkingTranscriptionalActivation2014}, Creative Commons license.}
    \label{F2014}
\end{figure}

Five major modifications are made in F2014 (Fig.~\ref{F2014}). First, the dynamics of the EC are remodeled in view of the recovery of clock function in \textit{elf4} mutants via ELF3 overexpression. This had previously been explained by the ability of all the components of the EC to bind to the PRR9 promoter \cite{herreroEARLYFLOWERING4Recruitment2012}.
In F2014 it is proposed that ELF4 enhances the effects of ELF3 through the formation of the ELF3-ELF4 complex, and consequently ELF3 can play this role, although with reduced efficacy. This way, the activity of the EC is modeled as rate-limited by LUX and NOX on one hand and ELF3-ELF4 on the other; the EC itself no longer has its own variable in the differential equations.

The addition of NOX, a close homologue of LUX, constitutes the second modification. NOX is assumed to act in parallel with LUX, serving as a transcriptional repressor in the EC. This addition enhances the model's accuracy, particularly in predicting peak timing under constant light conditions.

The third modification involves the inhibitor wave. Earlier models describe PRR activation sequentially, such that PRR9 activates PRR7, which then activates PRR5. This contradicts experiments wherein loss of function of one of the PRRs leaves the others unaffected. Instead, similarly to the observations that prompted the changes TOC1 underwent in P2012, later PRRs are found to repress the earlier ones. 

The fourth adjustment reassesses the role of CCA1 and LHY --now modeled independently. In previous models they repress all their targets except from PRR9 and PRR7. However, the inclusion of PRR9 activation by CCA1 and LHY does not improve the fit of the model, so F2014 neither requires nor rules out activation of PRR9 by CCA1 and LHY.

The fifth and final modification involves the introduction of the transcription factor RVE8. Activation by RVE8 is shown to be dispensable for clock function, as ascertained by the null effect on the cost of refitting the parameters of the model without including this component. Its inclusion is motivated mainly to explain the effects of the \textit{rve} mutants.

In parallel to the development of the other models of the expansion phase, Kim \textit{et al.} published a model that studies how the role of GI depends on its location inside the cell \cite{kimBalancedNucleocytosolicPartitioning2013}. Nuclear GI acts as a positive regulator of LHY expression, while cytosolic GI negatively regulates LHY. LHY in turn inhibits cytosolic GI. These interactions define a spatial network motif, namely a type-3 incoherent feedforward loop \cite{manganStructureFunctionFeedforward2003} between nuclear GI, cytosolic GI and LHY. The GI loop achieves the same dynamical capabilities of a three-component loop with only two molecular components, separated in different compartments. The system shows generation of strong pulses \cite{manganStructureFunctionFeedforward2003} for which balance of nuclear and cytosolic GI is shown to be crucial. Additionally, alternative modeling of the network using three distinct components shows that the two-component spatial feed forward loop is more robust against both internal and external noise. In summary, partitioning of the components of a gene regulatory network among intracellular compartments extends its regulatory capabilities, maintaining the number of components involved.

\section{Recent development of \textit{Arabidopsis} circadian clock models: Reduction phase}

The reduction phase refines the more complex models from the expansion phase, focusing on simplifying mathematics while maintaining the fundamental mechanisms of the clock. This process not only streamlines the models, but also enables a more thorough examination of their dynamic characteristics.
In addition to its mathematical elegance, model reduction is technically motivated, as exemplified by the \textit{sloppy model} framework \cite{gutenkunst2007universally,transtrumSloppy2015}. This approach indicates that when a model's complexity exceeds the available data's capacity to define its parameters, there are systematic procedures to reduce model size. This process consolidates less-critical processes into a broader, effective dynamic. Hence, leveraging biological insights to formulate simplified models from the outset represents a judicious strategy.

\begin{figure}[!t]
    \centering
    \includegraphics[width=0.8\linewidth]{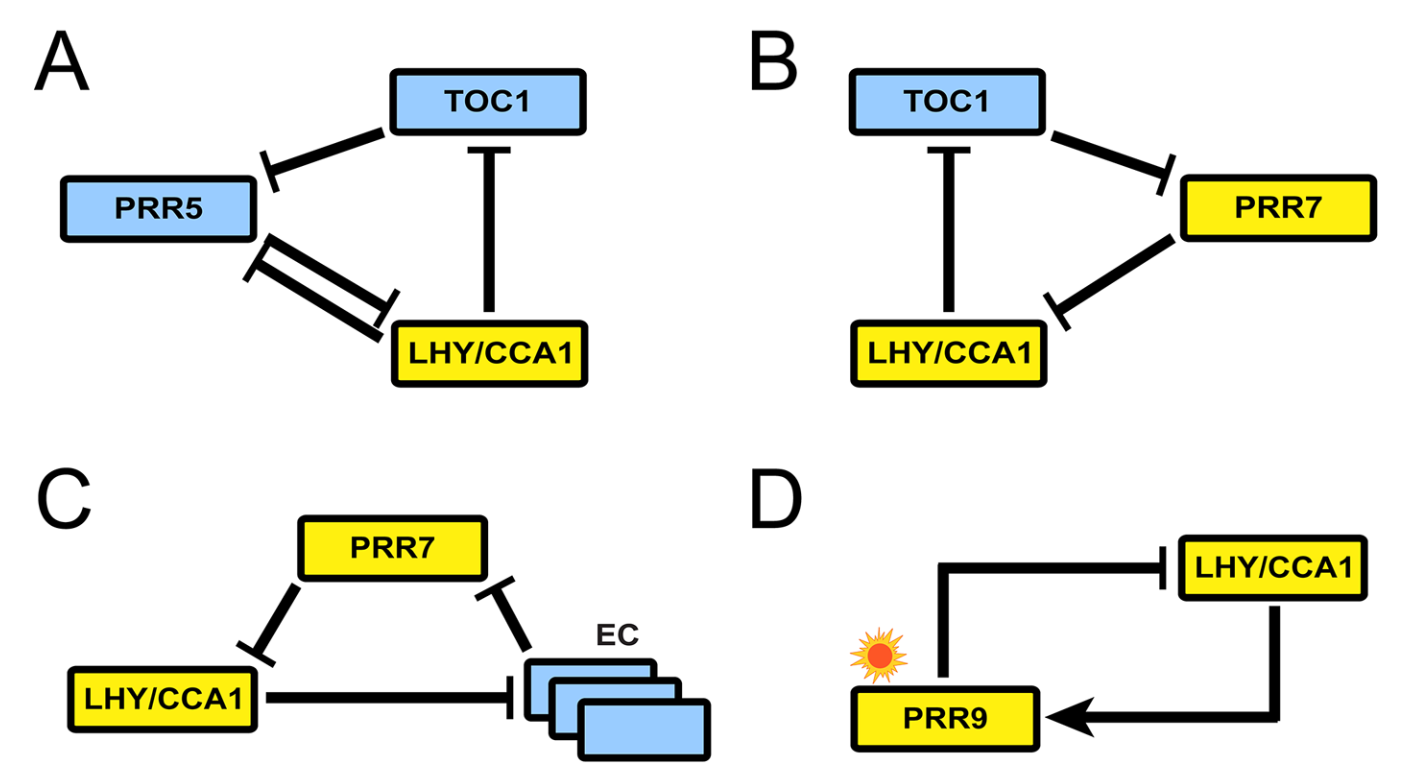}
    \caption{Structure of the kernel of the circadian clock in MF2015. The kernel consists of four loops, all containing at least one PRR and sharing LHY/CCA1. (A) Loop A, a repressilator involving TOC1, PRR5 and LHY/CCA1. (B) Loop B, a repressilator involving TOC1, PRR7 and LHY/CCA1. (C) Loop C, a repressilator involving the EC, PRR7 and LHY/CCA1. And (D) Loop D, a negative feedback loop between PRR9 and LHY/CCA1. Reproduced from \cite{fooKernelArchitectureGenetic2016}, Creative Commons license.}
    \label{MF2015}
\end{figure}

The reduction phase starts with two models published (coincidentally) in the same week of 2016. The first one by Foo \textit{et al.}, MF2015 \cite{fooKernelArchitectureGenetic2016}, introduces the notion of \textit{kernel} to the study of the plant molecular oscillator: a set of minimal regulatory modules that include all molecular components in the system and only the interactions necessary to generate the behavior observed in the wild type. This concept was first introduced in the context of evolutionary developmental biology \cite{davidsonGeneRegulatoryNetworks2006} and has also been defined as the minimal set of nodes inside a gene regulatory network whose regulation ensures the achievement of any attractor state given an initial network state \cite{kimDiscoveryKernelControlling2013}. In the case of circadian clocks, of course, the attractor will be a limit cycle. 
Foo and collaborators base their analysis on an updated version of P2012, that includes recent experimental findings and removes outdated interactions. Using a heuristic approach, they find that the kernel of the \textit{Arabidopsis} clock harbors four negative feedback loops, labelled A to D (Fig.~\ref{MF2015}). All of them contain at least one PRR and are interlocked by sharing LHY/CCA1. Structurally, loops A to C are repressilators, and loop D is a negative feedback loop. 

Loop I involves, besides the repressilator structure between LHY/CCA1, TOC1 and PRR5, the inhibition of PRR5 by LHY/CCA1, in a direction opposite to the repressilator's.
This regulatory motif was later named the AC-DC circuit \cite{perez2018combining}, and is able to display rich dynamical behavior.
This inhibition slows down the oscillation of the loop, an effect that extends to the whole clock because of the interconnection with the rest of the system. Simulations of the model show that a 20\% increase in PRR5 inhibition by LHY/CCA1 lengthens the period by 3.3h, and decreasing it by the same amount shortens it by 2.9h. Note that loops I and II are not described by previous models, while loop III echoes the repressilator in P2012.

Foo and collaborators use the kernel decomposition to explore whether any loop is critical to support the oscillations. Separate modeling of each module shows that only loop IV is unable to sustain autonomous oscillations. Meanwhile, loop I can make up for the loss of loops II and III under certain conditions, and loops II and III can buffer the loss of each other. Thus loop IV has a relatively weak role in free running rhythm despite it being the only entrainable one. Finally, the kernel decomposition allows the functional study of inhibitory interactions, which are shown to create cuspidate oscillatory profiles, potentially in service of accuracy in the timing of physiological events.

\begin{figure}[!t]
    \centering
    \includegraphics[width=0.5\linewidth]{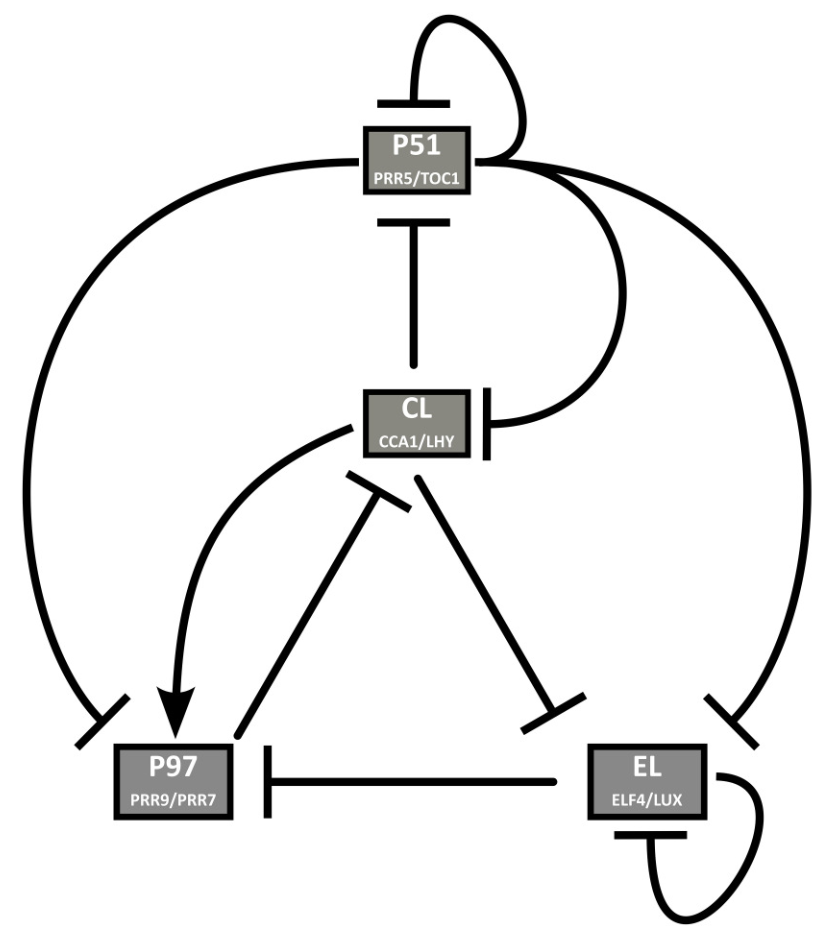}
    \caption{Structure of the circadian clock in DC2016. Several pairs of genes are combined, thus considerably reducing the complexity of the model. To the previous merging of CCA1/LHY (CL) and PRR9/7 (P97), they add PRR5/TOC1 (P51) and ELF4/LUX (EL). Reproduced from \cite{decaluweCompactModelComplex2016}, Creative Commons license.}
    \label{DC2016}
\end{figure}

The second model of the reduction phase was published by De Caluwé and collaborators, DC2016 \cite{decaluweCompactModelComplex2016}. Their approach is to merge genes that behave similarly in the network, to reduce the number of equations. The foregoing networks are reduced to four pairs of genes, in view of the large number of clock components and the largely redundant roles some of them  have. This way, several sets of almost identical equations are eliminated. In previous models, gene pairs such as LHY and CCA1 (in all models but F2014) or PRR7 and PRR9 (L2006, P2010) are considered single entities. DC2016 applies this strategy to two more pairs of genes, each of which have similar expression profiles, regulators, targets, and defects in loss-of-function mutant lines. The resulting network comprises nine ODEs, representing four nodes: P51 (PRR5/TOC1), CL (CCA1/LHY), P97 (PRR9/PRR7) and EL (ELF4/LUX) (Fig.~\ref{DC2016}). DC2016 retains key qualitative features of the circadian clock, and is also able to reproduce how the clock affects hypocotyl growth. Thanks to its simplicity, DC2016 has been so far the ``standard model'' used to study the link between the circadian system and physiological and developmental processes, and properties of the clock such as temperature compensation and entrainment. Some of these studies show the bidirectional relationship between the clock and the processes under its control, such as magnesium metabolism \cite{demeloMagnesiumMaintainsLength2021}, and the role of photosynthetic entrainment \cite{oharaGeneRegulatoryNetwork2018a, oharaPhotosyntheticEntrainmentCircadian2017}. DC2016 has also been used for stochastic modeling of the clock \cite{zhangStochasticSimulationModel2021}.

\begin{figure}[!t]
    \centering
    \includegraphics[width=0.8\linewidth]{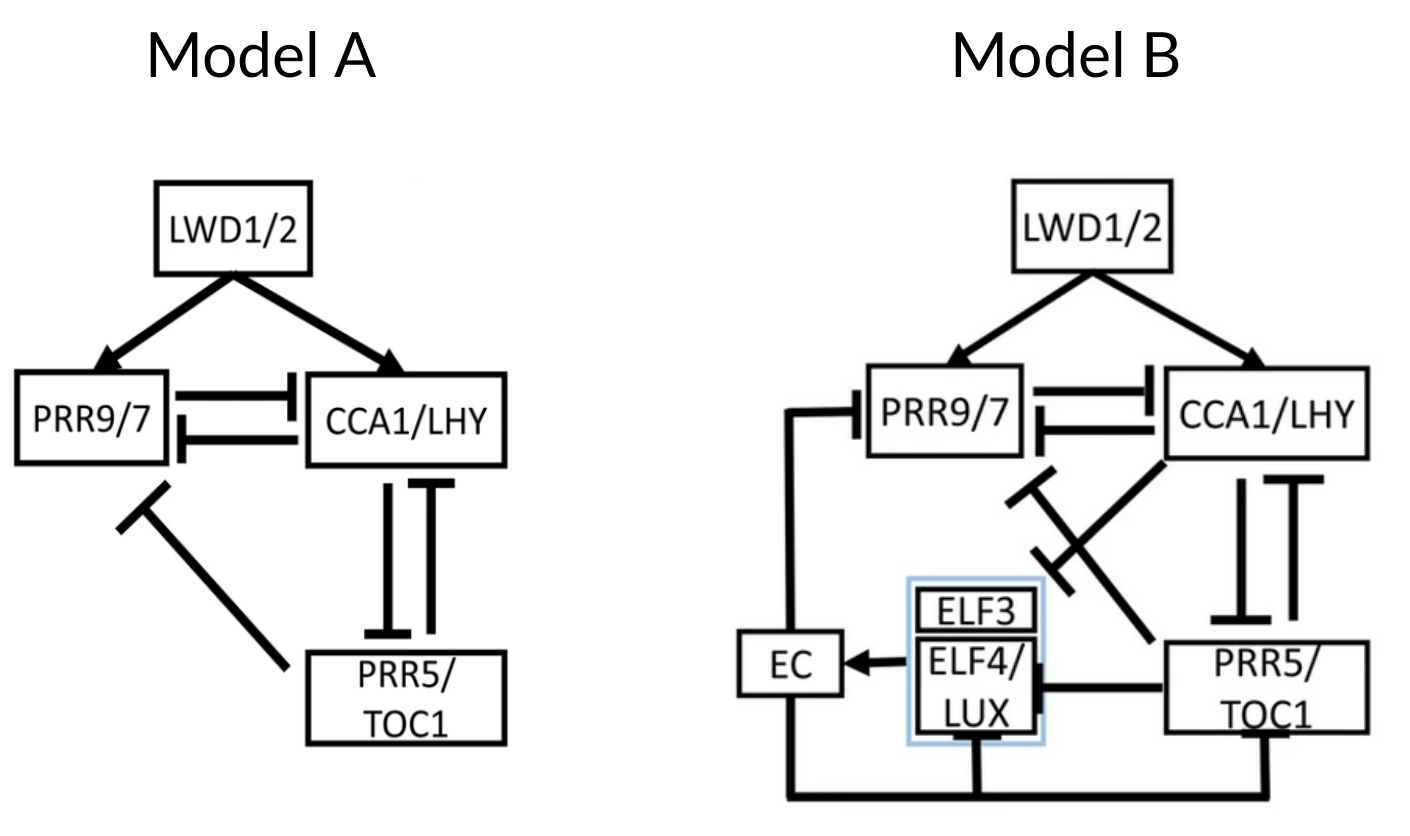}
    \caption{Structure of the core of the circadian clock in J2018. (Left) Model A is proposed as a core of the circadian clock, including the merged genes CCA1/LHY, PRR5/TOC1 and PRR9/7. As a novelty, the model includes activation by LWD 1/2. (Right) Model B adds the Evening Complex and its components with their respective interactions. This second model better captures the overall behavior of the clock. Modified from \cite{joanitoIncoherentFeedforwardLoop2018}, Creative Commons license.}
    \label{J2018}
\end{figure}

The next model in the reduction phase by Joanito \textit{et al.}, henceforth J2018 \cite{joanitoIncoherentFeedforwardLoop2018}, decomposes the clock network into compact regulatory modules, providing insights into the robustness and hysteretic behavior of the network. Using the merged gene structure of DC2016, they build a series of simplified models of increasing complexity and compare them, finding that the best description of the available data for both the wild type and several mutants is what they call ``Model A'' (Fig.~\ref{J2018}, left). This model is very similar to DC2016 without the EL component, but it only includes inhibitory interactions between CCA1/LHY, TOC1/PRR5 and PRR9/7. As a novelty, J2018 includes positive activation of CCA1/LHY and PRR9/7 by \textit{LIGHT-REGULATED WD1} and \textit{2} (LWD1/2).

Next, Joanito and collaborators explore the different network motifs present in Model A and their functional consequences. They find a repressilator between CCA1/LHY, PRR5/TOC1 and PRR9/7 (also described in DC2016) that must drive the oscillation of the whole clock, since Model A contains no other negative feedback loop nor self-oscillating components. Also, Model A contains a positive feedback loop consisting of a double inhibition between CCA1/LHY and PRR5/TOC1. This results in a toggle switch with bistability and hysteresis, where either CCA1/LHY is high and PRR5/TOC1 is low, or the reverse, depending on the expression of PRR9/7, which acts as the switch. Model A contains another positive feedback loop made of negative interactions between CCA1/LHY and PRR9/7 which does not result in a bistable behavior. Finally, Model A also contains three type-2 incoherent feed-forward loops \cite{manganStructureFunctionFeedforward2003} between the three components of the model, resulting in a pulse-like expression of PRR9/7 and CCA1/LHY. 

J2018 includes an additional model, named ``Model B'' (Fig.~\ref{J2018}, right). This is an extension of Model A that includes the EC and its components, ELF3 and ELF4/LUX, with their respective regulatory interactions. The presence of the EC adds another repressilator to the system (also present in DC2016), which results in a stronger toggle switch between CCA1/LHY and PRR5/TOC1. Additionally, the EC acts as a nighttime switch for the bistable behavior. Model B thus shows a complete alternating behavior between the two stable states: at dawn, CCA1/LHY is high and PRR5/TOC1 is low. This results in a pulse-like expression of PRR9/7 peaking near noon, which switches the system to the other bistable state at dusk: low CCA1/LHY and high PRR5/TOC1. Low CCA1/LHY results in EC accumulation, which decreases PRR5/TOC1 and eventually switches the system to the other bistable state near dawn. The authors repeat their analyses in P2012 and F2014, observing bistability in both of them, with some caveats: EC does not result in bistability in P2012, and only PRR7 (but not PRR9) shows pulse-like expression in F2014. These results highlight the usefulness of simplified models in studying dynamical behaviors.

Further simplification of the models has been achieved using distributed delays. Using this approach, the number of unknown parameters can be reduced by introducing distributed time delays to substitute interactions whose main dynamical value is delaying the timing of other processes. Distributed delay representations have been applied to L2005a, MF2015 and DC2016 \cite{tokudaReducingComplexityMathematical2019}, retaining the key dynamical properties of these models while significantly reducing the number of parameters.

\section{Current challenges and overview}

\subsection{Spatial models}
The models reviewed so far are mean-field approximations of the molecular behavior across different tissues and plant parts. However, it has been shown that the plant circadian clock has spatially-specific functions \cite{davisSpatiallySpecificMechanisms2022} and consequently its properties ---such as phase, period and entrainability--- and mechanisms are spatially heterogeneous \cite{nohalesSpatialOrganizationCoordination2021}. 
To effectively incorporate circadian clock models into the broader context of physiology and development, two primary considerations arise. Firstly, the establishment of phase differences among various clocks, which, based on empirical evidence, can arise from differential clock gene expression, variations in molecular clock structures, or differences in signal processing. Secondly, the nature of inter-clock coupling, with proposed mechanisms including both local and long distance of clock components
\cite{nohalesSpatialOrganizationCoordination2021, gouldCoordinationRobustSingle2018}.

\begin{figure}[!t]
    \centering
    \includegraphics[width=0.5\linewidth]{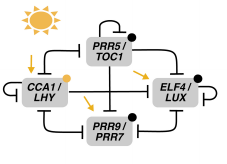}
    \caption{Structure of the circadian clock in G2022. The model presents an updated version of DC2016. The novelty here is the model of plant structure in hundreds of compartments, each of which has its own set of ODEs. Modified from \cite{greenwoodSpatialModelPlant2022}, Creative Commons license.}
    \label{G2022}
\end{figure}

The spatial model constructed by Greenwood \textit{et al}., G2022 \cite{greenwoodSpatialModelPlant2022}, tries to address these challenges. For the construction of G2022, a conserved clock architecture is assumed: an updated version of DC2016 wherein CL represses P97, CL represses its own transcription and P51 degradation rates in the dark are increased (Fig.~\ref{G2022}). This network is instantiated in each of the approximately 800 cells constituting a simplified seedling model, segmented in four regions of varying light sensitivities: cotyledon, hypocotyl, root and root tip. Cell-to-cell coupling is modeled through sharing of CCA1/LHY mRNA. This mechanism, together with differential light sensitivities, generates the two travelling waves of gene expression ---up and down the root--- that are experimentally observed \cite{gouldCoordinationRobustSingle2018}. Also, local coupling has a stabilizing effect on individual clock rhythms, buffering noise while maintaining the between-region phase differences.
On the other hand, long distance coupling is modeled by the apportioning of ELF4 from shoots to roots, shown to be non-essential to produce the spatial waves of gene expression. Instead, the model suggests it matches clock periods between different organs.

G2022 is a step towards understanding the plant circadian clock not as a single gene regulatory network embedded in every cell, but as a system spanning the whole plant. It is a separable system not just because of its components ---which are differentially expressed throughout the plant tissues--- but because of what it does ---its \textit{activity-function} \cite{jaegerDynamicalModulesMetabolism2021}, namely the proper timing of physiological processes. It also illustrates how bottom-up emergence (molecular oscillations) is sculpted by top-down constraints (tissue-dependent input sensitivities) \cite{juarreroContextChangesEverything2023}.

\subsection{Effects of light on the clock}

The models presented so far barely deal with the effect of light on the clock. Plants have a wide array of photoreceptors \cite{galvao2015sensing} detecting wavelengths from around 280 to 750 nm, that is, from UV-B to far-red light. These different photoreceptors translate the light signals into molecular actions, in many cases mediated through changes in the clock. In what follows, we discuss works that present mathematical models integrating light intensity and quality (through the inclusion of diverse photorreceptors) with clock models.

\begin{figure}[!t]
    \centering
    \includegraphics[width=0.8\linewidth]{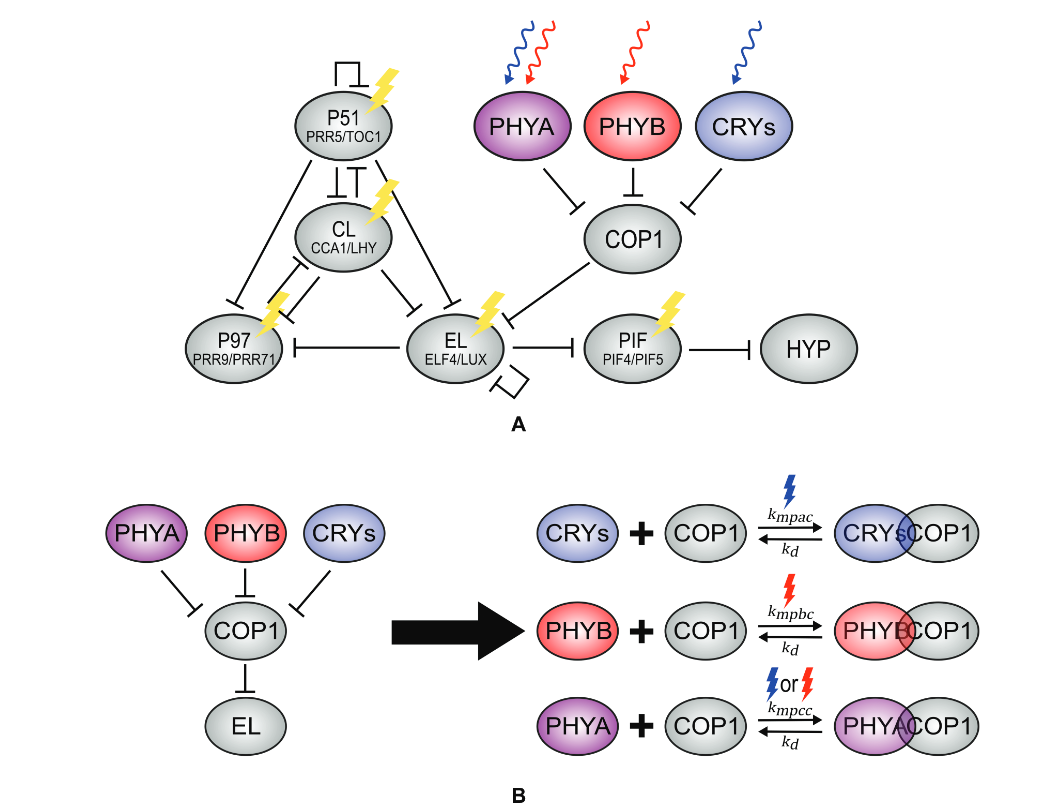}
    \caption{Structure of the circadian clock in Pa2022a. The model includes the perception of light by different photoreceptos (A) as well as their repression of COP1. The model also incorporates competitive binding between COP1 and phyA, phyB and cry (B). Reproduced from \cite{payModellingPlantCircadian2022}, Creative Commons license.}
    \label{Pa2022}
\end{figure}

Ohara and collaborators \cite{oharaExtendedMathematicalModel2015}, O2015, first introduced the effect of light on the clock through a modification of L2005a. They argue that this model captures the core functioning of the clock despite its simplicity. Their modification consists on replacing the light-sensitive protein P of L2005a by including equations for several photoreceptors: \textit{phytochromes A} and \textit{B} (phyA and phyB), responsible for detecting red light, and \textit{chryptochrome 1} (cry1), responsible for detecting blue light. The molecular concentrations of these photoreceptors are then translated into an ``effective light'' equation that alters the expression of LHY. O2015 is able to capture the phase response curves of the circadian clock after different pulses of light. In particular, O2015 recovers the atypical phase response curve obtained with a pulse of darkness, in contrast with L2005a or P2011, which predicted an erroneous response.

O2015 has recently been updated by Pay and collaborators \cite{payModellingPlantCircadian2022}, Pa2022a, who have applied the light functions of O2015 to a modified version of DC2016, in order to study the effect of light on hypocotyl growth. DC2016 is modified so that the activation of P97 by CL is changed to repression, as in F2014. Pa2022a also includes the competitive binding of COP1 to all photoreceptors, which inactivates them, as well as COP1's repression of EL and the photoreceptors' repression of COP1 (Figure \ref{Pa2022}), ending up with a model of 18 ODEs and 66 parameters. The model is able to qualitatively fit both phase response curves and hypocotyl growth measures under several light environments. 

More recently, Pay \textit{et al.} \cite{payExtendedPlantCircadian2022}, Pa2022b, have extended their model in order to study the effect of light on flowering. Notably, this new model includes GI as a regulator of flowering, but downstream of the clock. That is, GI has no effect on the circadian clock in this model.

Following an alternative route to O2015, Huang \textit{et al.} \cite{huang2022red} have studied the effect of red light on the circadian clock, focusing on the expression patterns of CCA1 through the action of phyB. In order to build their model, they use a modified version of DC2016, which they complement with the addition of a light-sensitive equation for ELF3 to make the model suitable for red-light environments. Fitting their model to experimental expression time series of CCA1, they show that a red-light pulse causes a phase-shift in CCA1 expression. They also predict that red light can reset both the phase and the period of the clock, and study how different red-light dynamic environments alter the parameters of the model, and thus the dynamical properties of the clock.

De Caluwé and collaborators \cite{decaluweModelingPhotoperiodicEntrainment2017} use a modified version of DC2016 wherein all the clock genes are light-regulated to study the mechanistic basis of entrainment of the clock to changing day lengths. The clock shows diverse behaviors upon exposure to non-24 hour light cycles, from frequency demultiplication to chaos. Studying these responses by following Schmal's theoretical entrainment paradigm \cite{schmalTheoreticalStudySeasonality2015}, they assess the theoretical entrainment range of the clock.

They discuss the change of the limit cycle behind the clock oscillations as daylength increases from total darkness to continuous light, and find parameter ranges where the behavior of the clock becomes chaotic.

\subsection{Effects of temperature on the clock}

The \textit{Arabidopsis} circadian clock shows ``temperature compensation'', that is, it is able to maintain a stable period in a broad range of temperatures. Some models have studied the molecular and regulatory basis of this effect. 

Gould \textit{et al.} \cite{gouldNetworkBalanceCRY2013} study the molecular basis of temperature compensation through a combination of experiments and modeling. They show a combined effect of temperature and light quality on temperature compensation. Thus, the clock period is shortened at higher temperatures under white and blue light, and this effect is dependent on cry. Both phyA and phyB also had nonlinear effects on the period at different temperatures, although the effects were small in comparison with those of cry. They also show that the expression of clock-related genes is almost constant over a range of temperatures, but this is also dependent on the presence of cry. In order to study these effects in detail, they introduce temperature dependence into P2010, resulting in a new model, G02013, which is able to capture their experimental findings and which predicts a temperature-dependent change in LHY, a result which was validated experimentally. Notably, their results show that the most important element in temperature compensation is light quality, most importantly cryptochromes through regulating multiple light and temperature-dependent parameters. 

More recently, Avello \textit{et al.} \cite{avelloHeatClockEntrainment2019} introduce temperature dependence through Arrhenius equations into DC2016.
Although useful to build qualitative insight into temperature effects, such a dependence has to be treated with caution, as for plants there is limited thermodynamic data available, such as activation energies, that would constrain an Arrhenius model.
Their model, A2019, is able to reproduce both temperature entrainment and compensation and point to the dependence on temperature of degradation rates as a driving force behind both processes across a range of temperatures. 

In a more recent work, Avello and collaborators \cite{avelloTemperatureRobustnessArabidopsis2021} introduce temperature dependence through Arrhenius equations in several of the models reviewed here: L2005a, L2005b, L2006, P2010, P2011, P2012, F2014 and DC2016. Following the approach by Joanito \textit{et al.} \cite{joanitoIncoherentFeedforwardLoop2018}, they analyze the structural patterns of the circadian clock that are behind temperature compensation. By listing all network motifs found in these clock models and studying how the period changes wih temperature in each of them, Avello and collaborators are able to conclude that no single network motif is responsible for temperature compensation. Rather, they find that temperature compensation is a global property of the network, where the presence of autoregulation combined with three-node feedback loops, plus a dominance of inhibitory interactions, results in the observed effect. Notably, it was the presence of these three factors that resulted in temperature compensation. This work highlights the use of minimal models as a way to better understand dynamical effects.

\section{Conclusions}

In the last twenty years, the structure of the clock has been modified and updated as new components and interactions were discovered. In fact, new interactions are continuously being discovered, see for instance \cite{yuan2024complex} where new interactions between ELF3 and PRR7 and PRR9 are described. While in a first phase mathematical models sought to incorporate as many components and interactions as possible, the last ten years have seen a shift in focus towards minimal models that can be better explored in terms of dynamical properties. In particular, DC2016 has become a golden standard in the field, being used in many subsequent works to study dynamical properties of the clock, introduce tissue differences, or to explore the effect of light and temperature.
These recent studies underscore the prevailing trend in the field: experimental testing and mathematical modeling of clock properties across a range of environmental conditions, with an emphasis on intra-plant heterogeneity. The focus is on elucidating underlying mechanisms rather than detailed molecular interactions. This effort aims not only to understand the clock itself, but also to explore its influence on clock-dependent phenotypes such as flowering and growth.

The strengths and weaknesses of both the expansion and reduction phases of modeling are evident in their respective contributions to our understanding of the \textit{Arabidopsis} circadian clock. Expansion phase models excel in their attention to molecular detail and their capability to make specific predictions for individual genes, providing a comprehensive representation of the clock's regulatory network. However, these models face significant challenges, particularly in the difficulty of unequivocally fitting parameters due to their mathematical complexity and the sheer number of variables involved.

On the other hand, reduction phase models offer notable advantages in terms of mathematical simplicity, allowing for parameters that are better constrained by experimental data. This simplicity enables a clearer understanding of the essential dynamical properties of the clock. Nonetheless, these models also have their limitations, as their parameters are often effective rather than mechanistic, making it challenging to make precise predictions for the dynamics of specific genes.

Looking ahead, future directions in this field should focus on leveraging the strengths of both modeling approaches. Reduced models, which can be better constrained by data, should serve as foundational templates for developing more complex models. These more intricate models should be capable of being reduced back to the well-tested reduced models, ensuring that they offer reliable insights while also providing detailed predictions for genes or variables of specific interest. This iterative approach will enhance our ability to balance molecular detail with practical simplicity, ultimately leading to a deeper understanding of the circadian clock's mechanisms and its broader biological implications.

\section{Author Contributions}
L.H., P.C., and S.A. contributed to the conception and design of the work. L.H. drafted the manuscript. L.H., P.C., and S.A. edited and revised the manuscript critically for important intellectual content. {All authors} have read and agreed to the published version of the manuscript.

\section{Author Competing Interests}
The authors declare no conflict of interest.

\section{Funding}
This research was funded by MCIN/AEI/10.13039/501100011033 and by ``ERDF/EU'' through grant PGE, grant numbers PID2022-142185NB-C21 and PID2022-142185NB-C22, to SA and PC.

\renewcommand\refname{References}
\begin{footnotesize}
\bibliographystyle{unsrt.bst} 

\end{footnotesize}
\newpage




\end{doublespacing}
\end{document}